\documentclass[a4paper, 10pt]{article}

\usepackage{times,xcolor,amsmath,amssymb,epsfig,graphics,graphicx}
\usepackage[normalem]{ulem}

\newcommand{\be}{\begin{eqnarray}}
\newcommand{\ee}{\end{eqnarray}}

\newcommand{\C}{\mathbb{C}}
\renewcommand{\bullet}{\centerdot}

\begin{document}

\begin{center}{\Large \bf Massless Rarita-Schwinger field from a divergenceless anti-symmetric-tensor spinor of pure spin-$3/2$}
\end{center}

\vspace{0.37cm}

\begin{center} {J. P.  EDWARDS$^1$ and M. KIRCHBACH$^2$}
\end{center}

\vspace{0.37cm}

\begin{center}
$^1$Instituto de F{\'{i}}sica y Matematicas, Universidad Michoacana de San Nicol\'as de Hidalgo, Avenida Francisco J.\ Mujica S/N, Edificio C-2, 
Ciudad Universitaria, C.P. 58030 Morelia, Michoacan,  M\'exico\\
{e-mail: jedwards@ifm.umich.mx }\\
{$^2$Instituto de F{\'{i}}sica, Universidad Auto\'noma de San Luis Potos{\'{i}},
Av. Manuel Nava 6, Zona Universitaria,
San Luis Potos{\'{i}}, S.L.P. 78290, M\'exico }
\\ e-mail: mariana@ifisica.uaslp.mx
\end{center}

\begin{flushleft}
{\bf Abstract:} We construct the Rarita-Schwinger  basis vectors, $U^\mu$,  spanning the direct product space, $U^\mu:=A^\mu\otimes u_M$,  of a massless four-vector, $ A^\mu $, with massless Majorana spinors, $u_M$,  together with the associated field-strength tensor, ${\mathcal T}^{\mu\nu}:=p^\mu U^\nu -p^\nu U^\mu$. The ${\mathcal T}^{\mu\nu}$ space is reducible and contains one 
massless subspace of a pure spin-$3/2$  $\in (3/2,0)\oplus (0,3/2)$. We show how to single out  the latter in a unique way by acting on ${\mathcal T}^{\mu\nu}$ with an earlier derived  momentum  independent projector, ${\mathcal P}^{(3/2,0)}$,  properly constructed from one of the Casimir operators of the algebra $so(1,3)$ of the  homogeneous Lorentz group. 
In this way it becomes possible to describe  the irreducible massless $(3/2,0)\oplus (0,3/2)$ carrier  space by means of the anti-symmetric-tensor of second rank with Majorana  spinor components, defined as $\left[ w^{(3/2,0) }\right]^{\mu\nu}:=\left[{\mathcal P}^{(3/2,0)}\right]^{\mu \nu}\,\,_{\gamma\delta}{\mathcal T}^{\gamma  \delta }$. The conclusion is  that the  $(3/2,0)\oplus (0,3/2)$ bi-vector spinor field can play the same role with respect to a $U^\mu$ gauge field as the bi-vector, $(1,0)\oplus (0,1)$, associated with the electromagnetic  field-strength  tensor, $F_{\mu\nu}$, 
plays for the  Maxwell gauge field, $A_\mu$. Correspondingly, we find the free electromagnetic field  equation, $p^\mu F_{\mu\nu}=0$, is paralleled by the free massless Rarita-Schwinger field equation, $p^\mu  \left[ w^{(3/2,0)}\right]_{\mu\nu}=0$,  supplemented by  the additional condition, 
$\gamma^\mu\gamma^\nu \left[ w^{(3/2,0)}\right]_{\mu \nu} =0$, a constraint that invokes the  Majorana sector.

\end{flushleft}


{\bf Pacs:}{Mathematics Subject Classification 2010: 22E70, 15A66, 35Q41, 81S05 }

\section{Introduction}
The spin-$3/2$ Rarita-Schwinger field, whether it be massive or massless,  appears in the  theory of  super-gravity  where it  defines the gravitino, the super-symmetric partner of the graviton, the gauge boson of a hypothesized fundamental gravitational interaction. Knowing its properties under Lorentz transformations  is indispensable for testing the predictive power of those field theories.  The gravitino is considered to transform according to the highest spin-$3/2$ of a neutral
four-vector spinor carrier space of the enveloping $sl(2,\C)$ algebra  of the homogeneous Lorentz group algebra, $so(1,3)$, its standard notation being, $(1/2,1/2)\otimes \left[(1/2,0)\oplus (0,1/2)\right]\sim A_\mu\otimes u_M$, with $A_\mu$ denoting one of the four basis vectors spanning the $(1/2,1/2)$ carrier space of that very same  algebra $sl(2,\C)$, and $u_M$ standing for a Majorana spinor. This particle  is supposed to obey the Rarita-Schwinger equation 
\cite{SAdler}--\cite{Deser}.
While for massive fields this equation is known to suffer several  inconsistencies, among them acausal propagation of the classical wave fronts within an electromagnetic background (Velo-Zwanzinger problem) and (the Johnson-Sudarshan problem) non-covariant equal-time commutators upon quantization (see \cite{Sorokin} for a review and references therein),  the massless field has been shown to be free of them. Specifically in  \cite{ADas} it has been shown that when considered as a supersymmetric partner to the graviton, the propagation of the massless gravitino is always causality and covariance  respecting. Later on, it was demonstrated that consistent quantum field theories for the massless gravitino can be constructed at both the classical and quantum levels,by employing  a combination of path-integral quantization with Hamiltonian constraint techniques  \cite{SAdler}, \cite{Dengiz}.
Instead, canonical quantization alone has turned out to be insufficient for fields transforming according to irreducible carrier spaces of the Lorentz group of the type, $(A/2,B/2)$ (with $A$, $B$ integer) characterized by multiple spins varying from  $|(A-B)|/2$ to $(A+B)/2$ because it is  prejudiced by Weinberg's theorem \cite{Wbrg}  according to which the helicity of the quantum state is limited to the lowest one, while the physical helicities 
have to correspond to the maximal allowed absolute value.
The  latter problem of quantization concerns primarily the four-vector used in the description of gauge fields. The canonical  quantization of a massless gauge field equipped by helicities 
$|\mp 1|$, can be achieved by means of a four-component field which, however,
does  not behave as a four-vector because it transforms inhomogeneously under Lorentz transformations,
\begin{eqnarray}
{\mathcal U}
(\Lambda)
A^\mu(x){\mathcal U}^{-1}(\Lambda) &=& 
\left( \Lambda ^{-1} \right)^\mu \,\, _\nu A^\nu(\Lambda x) +
i\frac{\partial }{\partial (\Lambda x)_\mu}\Omega (\Lambda x),
\label{zerohelcty}
\end{eqnarray}
where $\Lambda$ is a Lorentz transformation in space time, ${\mathcal U}(\Lambda)$ its representation on the space of the  $A^\mu(x)$ fields, and $\Omega (\Lambda x)$ is linear in the particle-annihilation, and anti--particle creation operators (for a detailed discussion see \cite{Kugo} and references therein). The way out is starting with a classical  Lagrangian whose kinetic term is  based upon the field-strength tensor, $F^{\mu\nu}$, known to transform as a single
 spin-$1$ and according to the $(1,0)\oplus (0,1)$ bi-vector carrier space of the Lorentz group.  This Lagrangian, denoted by ${\mathcal L}_0$ and  given by, 
\begin{eqnarray}
{\mathcal L}_0&=&-\frac{1}{4}F^{\mu\nu}F_{\mu\nu}, \quad 
F^{\mu\nu}=\partial^\nu A^\mu -\partial ^\mu A^\nu,
\label{lr_ginvt_lgr}
\end{eqnarray}
 is the only one that is both Lorentz-and gauge invariant, i.e. invariant under $A^\mu\to A^\mu -\partial^\mu \lambda $ transformations \cite{WeinbergChap}. In contrast, the Fermi Lagrangian ($-(1/2)(\partial _\nu A_\mu)(\partial ^\nu A^\mu)$) is only Lorentz- but not gauge invariant.  The aforementioned problems hint at the importance of the  pure spin carrier spaces of the Lorentz group for canonical  quantization.  The problems  of the four-vector quantization discussed above extend to the massless spin-$3/2$ Rarita-Schwinger field described by means of a four-vector spinor.  Also in this case, one may expect that 
expressing the kinetic term in the corresponding Lagrangian by means of  an anti-symmetric tensor-spinor, transforming according to  single spin-$3/2\in  (3/2,0)\oplus (0,3/2)$, may be useful to canonical quantization (as already pointed out in  \cite{Allcock}). \\

\noindent 
It is the goal of the present study to explicitly construct 
 the aforementioned $(3/2,0)\oplus (0,3/2)$ carrier space of the Lorentz group as a totally anti-symmetric tensor of second rank with Majorana spinor components,  and formulate the related classical Lagrangian in the hope that in this way we lay down the grounds  for a canonical constraint Hamilitonian quantization that could be generalized  to any spin. Motivated by the construction of massive four vectors in \cite{AK, ADK} we build them from direct products of Weyl (co-)spinors, extending that work to the massless case. We first construct the massless Dirac ($u$)  and Majorana ($u_M$) spinors,
and then also the massless four-vectors $A_\mu$. We then build the direct products of $F_{\mu\nu}$ with  Majorana spinors according to,
\begin{eqnarray}
{\mathcal T}^{\mu\nu}_{a}=p^\mu A^\nu\otimes \left[u_M \right]_a- p^\nu A^\mu \otimes \left[u_M \right]_a&=&p^\mu U^\nu_a -
p^\nu U^\mu_a, \nonumber\\
  U_a^\mu&:=&A^\mu\otimes \left[u_M\right]_a,
\label{gl1}
\end{eqnarray}
where $U^\mu_a$ denotes a Rarita-Schwinger four-vector spinor, while $\left[u_M\right]_a$ is  Majorana spinor.  In so doing, one
arrives at a 24 dimensional space spanned by totally anti-symmetric (with respect to boost transformations) massless tensors of second rank  with Majorana spinor components, i.e to 
$\left[ (1,0)\oplus (0,1)\right]\otimes \left[ (1/2,0)\oplus (0,1/2)\right]$. This carrier space splits into an eight dimensional irreducible sector of pure spin-$3/2$, 
a four-dimensional one of pure spin-$1/2$, and a 12 dimensional one of mixed  spins $1/2$ and $3/2$  according to
\begin{eqnarray}
\left[ (1,0)\oplus (0,1)\right]\otimes \left[ (1/2,0)\oplus (0,1/2)\right]&\Rightarrow&
 \left[
\left( \frac{3}{2},0\right)\oplus \left( 0,\frac{3}{2} \right) 
\right]
\oplus 
\left[ \left( \frac{1}{2},0 \right)\oplus \left( 0,\frac{1}{2}\right) \right] \nonumber\\
&\oplus& \left[\left(1,\frac{1}{2} \right)\oplus \left(\frac{1}{2},1 \right)  \right].
\label{rdct}
\end{eqnarray} 

The irreducible $(3/2,0)\oplus (0,3/2)$  building block in (\ref{rdct}) can be singled out upon application  to  ${\mathcal T}^{\mu\nu}_{a}$
 of a  momentum independent projector, ${\mathcal P}^{(3/2,0)}$,  earlier
properly constructed in \cite{AGK}  from one of the Casimir invariants of the inhomogeneous Lorentz group algebra as,
\begin{eqnarray}
\left[{\mathcal P}^{(3/2,0)}\right]_{\alpha \beta; \gamma \delta }&=&
\frac{1}{8} \left(\sigma_{\alpha\beta}\sigma_{\gamma\delta} +\sigma_{\gamma \delta}\sigma_{\alpha\beta}\right) -\frac{1}{12}\sigma_{\alpha\beta}\sigma_{\gamma\delta},
\label{F32_pryct}
\end{eqnarray}  
with $\sigma_{\mu\nu}$ standing for $\sigma_{\mu\nu}=i\left[\gamma_\mu,\gamma_\nu\right]/2$, where $\gamma_\mu$ are the Dirac matrices, whose Dirac indices, 
$\left[ \gamma_\alpha\right]_{ab}$ we suppressed for the sake of simplifying notation.
Then the
anti-symmetric tensor-spinor  constructs of the type,
\begin{eqnarray}
\left[{\mathcal P}^{(3/2,0)}\right]^{ \alpha \beta}\,\,_ { \gamma \delta }\left[{\mathcal  T}\right]^{ \gamma \delta }:=
\left[w^{(3/2,0)}\right]^{\alpha \beta},
\label{rdct_24}
\end{eqnarray}
transform under the  Lorentz group as the basis tensors of a  pure spin-$3/2$.  These tensors are divergence-less by construction,

\begin{eqnarray}
p_\alpha\left[w^{(3/2,0)}\right]^{\alpha \beta}&=&0, 
\label{div}
\end{eqnarray}
and obey in the Majorana sector the relation,
\begin{eqnarray}
\gamma_\alpha\gamma_\beta \left[w^{(3/2,0)}\right]^{\alpha \beta}&=&0. 
\label{gammas}
\end{eqnarray}
The equation (\ref{div}) qualifies the anti-symmetric tensor spinors,
$ \left[w^{(3/2,0)}\right]^{\alpha \beta}$  of pure spin-$3/2$ as field tensors for the massless Rarita-Schwinger gravitino.

In this way, an anti-symmetric tensor-spinor of pure spin-$3/2$  is furnished which is suitable for defining the kinetic term in the gravitino Lagrangian. This text provides the technical details needed for the realization of the concepts presented above and is organized as follows. In the next section a concise review of the fundamentals  of the $SL(2,\C)$ group, the universal covering of the homogeneous Lorentz group, is presented with the aim to reach in a transparent way the Weyl equations and their solutions, the Weyl spinors and co-spinors, which we then employ in section 3 in the construction of massless Majorana spinors,  and  massless four-vectors, thereby preparing the building blocks of the massless Rarita-Schwinger four-vector spinors.  We compare  the outcome for the massless four vectors  with Wigner's little group approach in section 4.  Section 5 is devoted  to our prime result, the construction of the  totally anti-symmetric tensor-spinor transforming as $(3/2,0)\oplus (0,3/2)$, together with some of its properties. The text closes  with a brief summary  section.

\section{The Weyl equations}
According to the contemporary understanding,  space and time are unified by transformations of the pseudo-orthogonal group $SO(1,3)$, the Lorentz group,  with the basis of its fundamental representation being given by a four-vector, $A_\mu$, with $\mu=0,1,2,3$, in standard notation \cite{Das}. The space spanned by the four-vectors is then the  Minkowski space of Einstein's special relativity. However, from a purely mathematical point of view, orthogonal and pseudo-orthogonal groups appear as factor groups of more basic groups, the so called spin-groups, and the fundamental representations of the former present themselves as tensor products of the fundamental representations of the latter. Specifically the homogeneous  Lorentz group is the factor group of $SL(2,\C)$ (the special linear group in a two dimensional complex space) with respect to its Abelian subgroup $Z_2$  
 (the center of the group) \cite{A2}, \cite{Rumer}. 

The $SL(2,\C)$ group has six generators defined by the Pauli matrices as, $\sigma_1/2$, $\sigma_2/2$, $\sigma_3/2$, and 
$i\sigma_1/2$, $i\sigma_2/2$, $i\sigma_3/2$, whose commutators,

\begin{eqnarray}
sl(2,\C):\quad \left[\frac{\sigma_i}{2},\frac{\sigma_j}{2}\right]=i\epsilon_{ijk}\frac{\sigma_k}{2}, &&
\left[\frac{i\sigma_i}{2},\frac{i\sigma_j}{2}\right]=-\epsilon_{ijk}\frac{i\sigma_k}{2},\nonumber\\
\left[\frac{\sigma_i}{2},\frac{i\sigma_j}{2}\right]&=&i\epsilon_{ijk}\frac{i\sigma_k}{2},
\label{sl2c}
\end{eqnarray} 
constitute the algebra, denoted by lower case letters as $sl(2,\C)$, of the $SL(2,\C)$ group. The commutators in (\ref{sl2c}) are a subset of the Clifford algebra of the lowest order.  This  group is known to have two non-equivalent fundamental representations, their respective bases being two dimensional vectors with complex components termed as spinors, $\zeta^\alpha$, and co-spinors, $\eta _{\stackrel{\bullet}{\beta}}$, with $\alpha=1,2$ and  $\stackrel{\bullet}{\beta}=\stackrel{\bullet}{1},\stackrel{\bullet}{2}$.  
The two spinors under discussion are related by charge conjugation according to \cite{A2}, \cite{Rumer},
\begin{eqnarray}
\left(
\begin{array}{c}
\zeta_{\stackrel{\bullet}{1}}\\
\zeta_{\stackrel{\bullet}{2}}
\end{array}
\right)&=&
C\left(
\begin{array}{c}
\zeta^1\\
\zeta^2
\end{array}\right)^\ast, \quad C=\left( \begin{array}{cc}
0&1\\
-1&0
\end{array}\right),
\end{eqnarray}
where $C$ is the metric tensor in spinor space, while ``$\ast$'' denotes complex conjugation. The $C$ matrix (equal to the two-dimensional Levi-Civita tensor $\epsilon_{\alpha\beta }=\epsilon^{\alpha\beta}$) serves to raise and lower indices in spinor/co-spinor space according to 
$\zeta_\alpha=\epsilon_{\alpha\beta}\zeta^\beta$, $\zeta_{\stackrel{\bullet}{\alpha}}=\epsilon_{\stackrel{\bullet}{\alpha}\stackrel{\bullet}{\beta}}\zeta^{\stackrel{\bullet}{\beta}}$   amounting to
\begin{eqnarray}
\zeta_1=\epsilon_{12}\zeta^2=\zeta^2, &\quad& \zeta_2=\epsilon_{21}\zeta^1=-\zeta^1,
\label{C_spnrs}\\
\eta_{\stackrel{\bullet}{1}}=\eta^{\stackrel{\bullet}{2}} &\quad & \eta_{\stackrel{\bullet}{2}}=-\eta^{\stackrel{\bullet}{1}}.
\label{Cinvrs_cspnrs}
\end{eqnarray}
The $SL(2,\C)$ transformations generated by $i\sigma_j/2$ act distinctly on the spinors and the co-spinors (also termed Van der Waerden spinors) and are given by the so-called right ({\bf R}) - and left ({\bf L})-handed boosts,
\begin{eqnarray}
{\bf R}:\quad\, e^{ -i{\vec p}\cdot\frac{i {\vec \sigma}}{2}} &=&\cosh \frac{\theta}{2} +\hat{\vec p}\cdot {\vec \sigma} \sinh \frac{\theta}{2}, \quad \theta =|{\vec p}|, \quad {\hat{\vec{p}} = \frac{\vec p}{\theta}}, \label{spnr}\\
{\bf L}:\qquad e^{ i{\vec p}\cdot\frac{ i{\vec \sigma}}{2}}&=&\cosh \frac{\theta}{2} -\hat{\vec p}\cdot {\vec \sigma} \sinh \frac{\theta}{2}, 
\label{cspnr}
\end{eqnarray} 
where ${\vec p}$ is the three momentum. The notion of ``left-handed'' / ``right-handed'' refers to the sign, positive versus negative, of the ${\vec \sigma}\cdot {\vec p}$ term.
{}For particles of mass $m$, one chooses
\begin{eqnarray}
\cosh \frac{\theta}{2}=\frac{E+m}{\sqrt{2m(E+m)}}, &\quad& \sinh\frac{\theta}{{2}} =\sqrt{\frac{m-E}{2m}}= \frac{|{\vec p}|}{\sqrt{2m(E+m)}}.
\label{cosh_sinh}
\end{eqnarray}
With $\sigma^{0} = \begin{pmatrix}1 & 0 \\ 0 & 1\end{pmatrix}$ the identity matrix, substitution of the last equations in (\ref{spnr}) and (\ref{cspnr}) amounts to the following expressions for the two boosts,
\begin{eqnarray}
 e^{ -i{\vec p}\cdot\frac{ i{\vec \sigma}}{2}} &=&\frac{1}{\sqrt{2m(E+m)}}\left((E+m)\sigma_0 +{\vec \sigma}\cdot {\vec p}\right),
\label{rspnr}\\
e^{ i{\vec p}\cdot\frac{ i {\vec \sigma}}{2}}\quad &=&\frac{1}{2m(E+m)}\left((E+m)\sigma_0 -{\vec \sigma}\cdot {\vec p}\right).
\label{lspnr}
\end{eqnarray}
The transformations generated by the $\sigma_i/2$ matrices alone,
$\exp (i\vec{p}\cdot {\vec \sigma})$, are the rotations constituting  an $SU(2)$ subgroup, and are the same for both types of spinors.

It can be shown that all the irreducible carrier spaces of the $sl(2,\C)$ algebra can be constructed from reducing all the possible direct products of $r$ spinors, and $n$ co-spinors with $r$ and $n$ taking all non-negative natural values. Specifically,  direct products of spinors and co-spinors give rise to the four-vectors defining the fundamental representation of the Lorentz group, $SO(1,3)$, which are understood in the above scheme as spinor--co-spinor tensors of rank one, i.e. $A_\mu\sim \zeta^\alpha \eta^{\stackrel{\bullet}{\beta}}$, the conventional representation (that is not unique)  being
\begin{eqnarray}
\left( \begin{array}{c}\zeta^1\\ \zeta^2\end{array}\right) \otimes \left(\begin{array}{c}
\eta^{\stackrel{\bullet}{1}}\\\eta^{\stackrel{\bullet}{2}}\end{array} \right)&=&
\left(\begin{array}{cc}
\zeta^{1}\eta^{\stackrel{\bullet}{1}}&\zeta^{1}\eta^{\stackrel{\bullet}{2}}\\
\zeta^{2}\eta^{\stackrel{\bullet}{1}}&\zeta^{2}\eta^{\stackrel{\bullet}{2}}
\end{array}\right)=
\left(\begin{array}{cc}
-\zeta^{1}\eta_{\stackrel{\bullet}{2}}&\zeta^{1}\eta_{\stackrel{\bullet}{1}}\\
-\zeta^{2}\eta_{\stackrel{\bullet}{2}}&\zeta^{2}\eta_{\stackrel{\bullet}{1}}
\end{array}\right)=
\left( \begin{array}{cc}
A_0+A_3&A_1-iA_2\\
A_1 +iA_2&A_0-A_3
\end{array}
\right).\nonumber\\
\label{Gl11}
\end{eqnarray}
Here, $A_0=A^0$ is the time-like component of $A_{\mu}$, while $A_1=-A_x$, $A_2=-A_y$, and $A_3=-A_z$ are the corresponding three space-like components. 
Along the prescription in (\ref{Gl11}), four-derivatives in spinor- and co-spinor spaces are defined as,
\begin{eqnarray}
i\partial^{\alpha \stackrel{\bullet}{\beta}}&=& 
i\left(
\begin{array}{cc}
\partial_0+\partial_3& \partial_1-i\partial_2\\
\partial_1+i\partial_2&\partial_0 -\partial_3
\end{array}
\right)=
i\partial^0\, \sigma_0 +i\sum_i\partial_i\sigma^i=
p^0\sigma_0 +\vec{p}\cdot {\vec \sigma},\nonumber\\
 p_\mu&=&i\partial_\mu =i\frac{\partial }{\partial x^\mu },
\label{Weylleft}\\
i\partial_{\alpha \stackrel{\bullet}{\beta}}&=& i\left(
\begin{array}{cc}
\partial^0+\partial^3& \partial^1+i\partial^2\\
\partial^1-i\partial^{2}&\partial^0 -\partial^3
\end{array}
\right)=
i\partial^0\big[\sigma_0\big]^{T} +\sum_ii\partial ^i \big[ \sigma^i\big]^T=
p^0\sigma_0 -{\vec p}\cdot {\vec \sigma}^T, \nonumber\\
p^\mu&=& {i}\partial^\mu=i\frac{\partial }{\partial x_\mu}.
\label{Weylright}
\end{eqnarray}
where the upper script $T$ stands for ``transpose'' {and we use metric signature $(+, -, -, -)$.}

Spinors and co-spinors then satisfy the following kinematic equations,
\begin{eqnarray}
i\partial^{\alpha \stackrel{\bullet}{\beta}}\eta_{\stackrel{\bullet}{\beta}}&=&m \zeta^\alpha,\\
i\partial_{\alpha \stackrel{\bullet}{\beta}}\zeta^\alpha &=&m \eta_{\stackrel{\bullet}{\beta}}\,,
\end{eqnarray}
where $m$ is a constant mass. Following this, dynamics is introduced in the standard way by gauging the derivatives. In the massless case of interest here, the above equations reduce to the so called right- ({\bf R}) and left-({\bf L}) handed 
Weyl equations \cite{Rumer},

\begin{eqnarray}
{\mathbf R}:\qquad\, \left(p^0\sigma_0 +{\vec p}\cdot {\vec\sigma}\right)\stackrel{\bullet}{\phi}&=&
\left(\begin{array}{cc} 
E+p_z&p_x-ip_y\\
p_x+ip_y&E-p_z
\end{array}\right)\left( \begin{array}{c}
\phi_{\stackrel{\bullet}{1}}\\
\phi_{\stackrel{\bullet}{2}}
\end{array}\right)=0,
\label{RWeyl}\\
{\mathbf L}:\quad \left(p^0\, \sigma_0 -{\vec p}\cdot {\vec\sigma}^T\right)^T\chi&=&\left(\begin{array}{cc} 
E-p_z&-(p_x-ip_y)\\
-(p_x+ip_y)&E+p_z
\end{array}\right)\left( \begin{array}{c}
\chi^1 \\
\chi^2
\end{array}\right)=0.
\label{Weyl}
\end{eqnarray}
The solutions to the right-handed Weyl equations are the Weyl co-spinors, while the solutions of the  respective left-handed equations are  Weyl spinors, here in turn  denoted by $\stackrel{\bullet}{\phi}$ and $\chi$, respectively.

\noindent It is straightforward to check that the following (so far not normalized) solutions for $\stackrel{\bullet}{\phi}$ and $\chi$ hold valid:

\begin{eqnarray}
\stackrel{\bullet}{\phi}=\left(\begin{array}{c}
-(p_x-ip_y)\\
E+p_z
\end{array}\right), &\quad& 
\chi =\left( \begin{array}{c}
E+p_z\\
p_x+ip_y
\end{array}\right).
\label{Weyl1} 
\end{eqnarray}
In setting $p_x=p_y=0$, and $E=p_z$ one sees that $\stackrel{\bullet}{\phi}$ corresponds to helicity $(-1/2)$, while the helicity of $\chi$ is $(+1/2)$. When the co-spinor $\stackrel{\bullet}{\phi}$ carries a spin that is anti-parallel to the $z$ axis (in which case the symbol $\downarrow $ is used), while the spin of the  $\chi$ spinor is oriented along it, $\uparrow$, then  two more Weyl equations, describing co-spinors with $\uparrow$ and spinors with $\downarrow$, 
can be  obtained from the above two by reversing the momentum as,  ${ p_z}\to -{p_z}$.  Under this change (\ref{Weylright}) becomes,
\begin{eqnarray}
 \left(\begin{array}{cc} 
E-p_z&(p_x-ip_y)\\
(p_x+ip_y)&E+p_z
\end{array}\right)\left( \begin{array}{c}
\tau_{\stackrel{\bullet}{1}}\\
\tau_{\stackrel{\bullet}{2}}
\end{array}\right)=0, \quad \stackrel{\bullet}{\tau}=
\left(
\begin{array}{c}
E+p_z\\
-(p_x+ip_y)
\end{array}
\right).
\label{Weyl2}
\end{eqnarray}
Now subjecting (\ref{Weylleft})  to the same change, results in
\begin{eqnarray}
\left(\begin{array}{cc} 
E+p_z&-(p_x-ip_y)\\
-(p_x+ip_y)&E-p_z
\end{array}\right)\left( \begin{array}{c}
\rho^1\\
\rho^2
\end{array}\right)=0, \quad \rho=\left(\begin{array}{c}
p_x-ip_y\\
E+p_z
\end{array} \right),
\label{Weyl3}
\end{eqnarray}
where the respective solutions have been denoted by $\stackrel{\bullet}{\tau}$, and $\rho$. 

It can be checked that the four spinors, $\stackrel{\bullet}{\phi}, \rho,\chi $, and $\stackrel{\bullet}{\tau}$ are orthogonal and they are all necessary as building blocks of Dirac's massless four-component $u$ and $v$ spinors. The four different direct products among the above  Weyl spinors with the co-spinors  will be shown in the following to provide the  building blocks of the massless four vector. Finally,  we wish to point out that the Weyl equations  can alternatively be derived also from the representation theory of the inhomogeneous Lorentz group with the aid of the Pauli-Lubanski pseudo-vector, a result reported in  \cite{Lanfear}, \cite{Suslov}.

\section{Massless Dirac- and  Majorana-spinors, and massless four-vectors from Weyl spinors}

\subsection{Dirac and Majorana  spinors}
The massless Weyl spinors and co-spinors in the above equations (\ref{Weyl1}), (\ref{Weyl2}), and  (\ref{Weyl3}), now with suitable normalization, can be employed in the construction of Dirac's massless $u_\pm $ spinors according to
\begin{eqnarray}
u_+ =\left( \begin{array}{c}
\chi\\
\stackrel{\bullet}{\tau}
\end{array}
\right)&=&\frac{1}{\sqrt{2p_z(E+p_z)}}\left(\begin{array}{c}
E+p_z\\
p_x+ip_y\\
E+p_z\\
-(p_x+ip_y)
\end{array} \right),\label{uup} \\
u_-=\left( \begin{array}{c}
\rho\\
\stackrel{\bullet}{\phi}
\end{array}\right)&=&\frac{1}{\sqrt{2p_z(E+p_z)}}\left(\begin{array}{c}
p_x-ip_y\\
E+p_z\\
-(p_x-ip_y)\\
E+p_z
\end{array} \right).
\label{u_d}
\end{eqnarray}
The corresponding $v_\pm $ spinors are then obtained from the $u_\pm $ spinors through  the relationship, $v_\pm =\gamma_5u_\pm $, where  $\gamma_5 = \textrm{diag}({\mathbf 1}_{2\times 2}, -{\mathbf 1}_{2\times 2} )$, while  ${\mathbf 1}_{2\times 2}$ stands for the two-dimensional unit matrix. In effect, one arrives at 
\begin{eqnarray}
v_+ =\left( \begin{array}{c}
\chi\\
-\stackrel{\bullet}{\tau}
\end{array}
\right)=\frac{1}{\sqrt{2p_z(E+p_z)}}\left(\begin{array}{c}
(E+p_z)\\
(p_x+ip_y)\\
-(E+p_z)\\
p_x+ip_y\\
\end{array} \right),\label{vup} \\
v_- =\left( \begin{array}{c}
\rho\\
-\stackrel{\bullet}{\phi}
\end{array}
\right)=\frac{1}{\sqrt{2p_z(E+p_z)}}\left(\begin{array}{c}
p_x-ip_y\\
E+p_z\\
p_x-ip_y\\
-(E+p_z)\\
\end{array} \right),\label{vd} 
\end{eqnarray}
It can be verified that the $u_\pm$ and $v_\pm $ spinors satisfy the massless Dirac equations. As an example, for  $u_+$ one finds,
\begin{eqnarray}
\left( \begin{array}{cccc}
0&0&E-p_z&p_x-ip_y\\
0&0&p_x+ip_y&E+p_z\\
E-p_z&-(p_x-ip_y)&0&0\\
-(p_x+ip_y)&E+p_z&0&0
\end{array}
\right) u_+=0.
\label{msslss_Drc}
\end{eqnarray}
In a similar way, the remaining equations can be worked out with the aid of the Weyl equations in (\ref{RWeyl})-(\ref{Weyl3}).
{}From the above spinors one now can obtain the Majorana spinors, henceforth denoted by $u_M$, and defined in the standard way as
\begin{align}
u_M^+&=\left( 
\begin{array}{c}
\chi\\
i\sigma^2\left[\chi \right]^*
\end{array}\right)
= \frac{1}{2p_z(E+p_z)}
\left(\begin{array}{c} 
E+p_z\\
p_x+ip_y\\
p_x-ip_y\\
-(E+p_z)
\end{array}
\right)\,=\left( 
\begin{array}{c}
\chi\\
-\stackrel{\bullet}{\phi}
\end{array}
\right),  \label{Mjor_12}\\
u_M^-&=\left( 
\begin{array}{c}
\rho\\
i\sigma^2\left[\rho \right]^*
\end{array}\right)
= \frac{1}{2p_z(E+p_z)}
\left(\begin{array}{c} 
p_x-ip_y\\
E+p_z\\
E+p_z\\
-(p_x+ip_y)
\end{array}
\right)=\left( 
\begin{array}{c}
\rho\\
\stackrel{\bullet}{\tau}
\end{array}
\right). \label{Mjor_mn12}
\end{align}
Correspondingly, the massless equation satisfied by, say, $u^+_M$, reads,
\begin{eqnarray}
\left( \begin{array}{cccc}
0&0&E+p_z&p_x-ip_y\\
0&0&p_x+ip_y&E-p_z\\
E-p_z&-(p_x-ip_y)&0&0\\
-(p_x+ip_y)&E+p_z&0&0
\end{array}
\right) u_M^+=0,
\label{msslss_Mjr}
\end{eqnarray}
and so forth.

\subsection{ Massless $(1/2,1/2)$ polarization vectors}
  
In the current section we consider massless four-vectors as direct products of  Weyl spinors and co-spinors, an approach inspired by \cite{AK}, \cite{ADK} where the  massive four-vectors have been described in terms of direct products of massive left- and right-handed spinors. {}For that purpose we begin by calculating $\chi\otimes \stackrel{\bullet}{\tau}$,
$\rho\,\otimes \stackrel{\bullet}{\phi}$, $\chi\otimes \stackrel{\bullet}{\phi} $ and  $\rho\,\otimes \stackrel{\bullet}{\tau}$. In making use of the equations (\ref{Weyl1})-(\ref{Weyl3}) we find the following expressions:
\begin{align}
w_1=\chi\,\otimes \stackrel{\bullet}{\tau}&=
\frac{1}{2p_z(E+p_z)}\left( \begin{array}{c} 
(E+p_z)^2\\
-(E+p_z)(p_x+ip_y)\\
(p_x+ip_y)(E+p_z)\\
-(p_x+ip_y)^2
\end{array}\right),\label{w1}\\
w_2=\frac{1}{{\sqrt{2}}}\left( \chi\,\otimes \stackrel{\bullet}{\phi} +\rho\,\otimes \stackrel{\bullet}{\tau}\right)
&=\frac{1}{2\sqrt{2}p_z(E+p_z)}\left(\begin{array}{c} 
0\\
(E+p_z)^2 -(p_x-ip_y)(p_x+ip_y)\\
(E+p_z)^2 -(p_x-ip_y)(p_x+ip_y)\\
0
\end{array}
\right),\nonumber\\\label{w2}\\
w_4=\frac{1}{{\sqrt{2}}}\left( \chi\otimes \stackrel{\bullet}{ \phi} -\rho\,\otimes \stackrel{\bullet}{\tau}\right)&=
\frac{1}{\sqrt{2}p_z }\left( \begin{array}{c}
-(p_x-ip_y)\\
E\\
-E\\
p_x+ip_y
\end{array}\right),\label{w4}\\
w_3=\rho\,\otimes \stackrel{\bullet}{\phi}&=\frac{1}{2p_z(E+p_z)}\left( 
\begin{array}{c}
-(p_x-ip_y)^2\\
(p_x-ip_y)(E+p_z)\\
-(E+p_z)(p_x-ip_y)\\
(E+p_z)^2
\end{array}
\right).\label{w3}
\end{align}
Next we introduce the  matrix $S$ defined  according to \cite{AK}, \cite{ADK}, \cite{Barut} as,
\begin{eqnarray}
S=\frac{1}{\sqrt{2}}\left( 
\begin{array}{cccc}
0&i&-i&0\\
-i&0&0&i\\
1&0&0&1\\
0&i&i&0
\end{array}
\right),
\label{Smatrix}
\end{eqnarray}
and subject the $w_i$ vectors (with $i=1,2,3,4$) from above to the transformation $Sw_i=A_i$. In so doing we find,

\begin{align}
A_1=Sw_1&=\frac{-i}{2\sqrt{2}p_z(E+p_z)}\left( 
\begin{array}{c}
2(E+p_z)(p_x+ip_y)\\
(E+p_z)^2+(p_x+ip_y)^2 \\
i\left[(E+p_z)^2-(p_x+ip_y)^2\right]\\
0
\end{array}
\right), \label{W1}\\
 A_3=Sw_3&=\frac{i}{2\sqrt{2}p_z(E+p_z)}\left( 
\begin{array}{c}
2(E+p_z)(p_x-ip_y)\\
(E+p_z)^2+(p_x-ip_y)^2\\
-i\left[(E+p_z)^2-(p_x-ip_y)^2\right]\\
0
\end{array}
\right),\label{W3}\\
A_4=Sw_4&=\frac{i}{p_z}\left( 
\begin{array}{c}
E\\
p_x\\
p_y\\
0
\end{array}
\right), \quad -iA_4^\alpha =\frac{p^\alpha}{p_z},\quad \frac{p^\alpha p_\alpha}{p_z^2} =\frac{E^2-p_x^2-p_y^2}{p_z^2}=1,\nonumber\\
\alpha &=0,1,2,\quad p^0=E, \quad p^1=p_x,\quad p^2=p_y, \quad p^3=p_z,
 \label{W44}
\end{align}

and 

\begin{eqnarray}
A_2=Sw_2&=&
i\frac{(E+p_z)^2 -p_x^2-p_y^2}{2 p_z(E+p_z)}
\left(\begin{array}{c}
0\\
0\\
0\\
1
\end{array}
\right)=i\left(\begin{array}{c}0\\0\\0\\1\end{array} \right)=ie_z, \quad E^2-{\mathbf p}^2=0.\label{W2_lngt} 
\end{eqnarray}
It can be shown that all the $A_i$ four-vectors are mutually orthogonal for $p^2=0$ according to the Minkowski metric, $G$, 
\begin{eqnarray}
{\bar A}_iA_j=\delta_{ij},&&  {\bar A}_j= A_j^\dagger G,\quad G=\mbox{diag}(1,-1,-1,-1),\label{ort13}
\end{eqnarray}
Also because $A_4$ is equal to the $SO(1,2)$ linear momentum (see the following section), its orthogonality to the remaining vectors means that the $A_1$, $A_2$, and $A_3$ polarization vectors can be considered as transverse  to the direction of propagation,  
\begin{eqnarray}
A_4 {\bar A}_j&=& \frac{i}{p_z}p^\rho \left[ {A_j}\right]_{\rho} =0, \quad j=1,2,3.
\label{trnsvrslt}
\end{eqnarray}
For the sake of what follows, it is important to notice that the vectors $A_1$ and $A_3$ are also divergenceless with respect to the massless four-momentum
$p^\mu p_\mu=0$ according to,
\begin{eqnarray}
p_\mu A_1^\mu =p_\nu  A_3^\nu=0, \quad p^\mu=\left( 
\begin{array}{c}
E\\
p_x\\p_y\\
p_z 
\end{array} \right), \quad \mbox{with} \quad E^2=p_x^2+p_y^2 +p_z^2.
\label{divA_1}
\end{eqnarray}
Notice that for the  $p_x=p_y=0$ kinematic, the $(-iA_4)$ vector becomes purely time-like
\begin{eqnarray}
-iA_4=\frac{1}{E}\left( \begin{array}{c} E\\0\\0\\0 \end{array}\right). 
\label{k_vector}
\end{eqnarray}
Since the $A_2$ vector is constant in any frame -- see (\ref{W2_lngt}) -- it also falls  into the trivial representation of $SO(1,2)$ and thus decouples from the  remaining three, it is no longer available to the construction of a massless vector, a reason for which the Weyl spinor approach describes massless momenta in an implicit way through the mass-shell condition on the light cone in (\ref{W2_lngt}). The Weyl spinor approach does not lead to any null  vectors.
More light on this issue will be shed in the  next section.

\subsection{Classification of the massless four-vectors by the Casimir invariant of the $so(1,2)$ algebra}
It is straightforward to verify that the $A_i$ vectors span bases of $SO(1,2)$ irreducible representations.
In order to see this, recall the $so(1,2)$ algebra given by
\begin{eqnarray}
so(1,2):\qquad \left[K_+,K_- \right]&=&-2L_z,\nonumber\\
\left[K_\pm, L_z \right]&=&\mp K_\pm ,\nonumber\\
K_\pm &=& K_x \pm iK_y.
\label{so12_algbr}
\end{eqnarray} 
The respective generators of  boosts along the $x$ and $y$ axes, $K_x$ and $K_y$, and the generator of rotation around the $z$ axis, $L_z$ ,
are represented by the following matrices,  

\begin{eqnarray}
L_z&=&\left(\begin{array}{cccc}
0&0&0&0\\
0&0&-i&0\\
0&i&0&0\\
0&0&0&0
\end{array} \right), \quad K_x=\left(\begin{array}{cccc}
0&i&0&0\\
i&0&0&0\\
0&0&0&0\\
0&0&0&0
\end{array} \right), \quad K_y=\left(\begin{array}{cccc}
0&0&i&0\\
0&0&0&0\\
i&0&0&0\\
0&0&0&0
\end{array} \right).
\label{so12_gnrtrs}
\end{eqnarray}
The $so(1,2)$ algebra has one Casimir invariant, ${\mathcal C}$, defined as
\begin{eqnarray}
{\mathcal C}=L_z^2-K_x^2-K_y^2,
\label{So12_Csmr}
\end{eqnarray}
whose action on the $A_i$ vectors is
\begin{eqnarray}
{\mathcal C}A_i=f(f+1) A_i=2 A_i \quad\mbox{with}\quad f=1 \quad \mbox{for}\quad i=1,3,4, \quad {\mathcal C}A_2=0,
\label{Csmr_act}
\end{eqnarray}
meaning that $A_1$, $A_3$, and $A_4$  provide the basis of   a three-dimensional  non-unitary  $SO(1,2)$ representation, while $ A_2$ defines  an $so(1,2)$ singlet.

The $so(1,2)$ Casimir invariant  in (\ref{So12_Csmr}) can be simultaneously diagonalized with any one of the three generators  in (\ref{so12_gnrtrs}) constituting the $so(1,2)$ algebra, for example with $L_z$,  arriving in this way to vectors that, next to the $f$-quantum number, are labelled also by the $L_z$ eigenvalues, denoted by $m$, according to
\begin{eqnarray}
{\mathcal C}\left|f,m \right> &=&f(f+1)\left|f,m\right>, \quad L_z\left|f,m\right>=m\left|f,m\right>, \quad |m|\leq f.
\label{so12_labels}
\end{eqnarray}
However, because $L_z$ is not an $ so(1,2)$ invariant, such a simultaneous diagonalization is not frame independent. However, in the particular frame,  $p_x=p_y=0$,  the standard textbook right-and left-handed circular polarization vectors, $\epsilon_{1,1}$, and $\epsilon _{1,-1}$ are recovered as,
\begin{eqnarray}
-iA_1\longrightarrow \left|1,1\right>&=& \epsilon_{1,1} =-\frac{1}{\sqrt{2}}\left(\begin{array}{c}0\\1\\i\\0\end{array} \right),
\label{eps1}\\
-iA_3\longrightarrow \left|1,-1\right> &=&\epsilon_{1,-1}=\frac{1}{\sqrt{2}}\left(\begin{array}{c}0\\1\\-i\\0\end{array} \right),\label{eos_1mn1}\\
-iA_4\longrightarrow \left|1,0\right>&=&\epsilon_{1,0} =\left(\begin{array}{c}1\\0\\0\\0 \end{array} \right)\equiv n,\label{epszero}
\end{eqnarray}
where $\epsilon_{1,0}\equiv n$ is a purely time-like unit vector. 
The operators $K_\pm=K_x\pm iK_y$ ladder in the basis of (\ref{eps1})-(\ref{epszero}) according to,
\begin{eqnarray}
K_+\epsilon_{1,1}=K_-\epsilon_{1,-1}=0,&\quad& K_+\epsilon_{1,-1}=K_-\epsilon_{1,1}=\sqrt{2}e^{i\pi}n, \quad K_\pm n =\mp \sqrt{2}\epsilon_{1,\pm 1}. 
\label{so12_lddrs}
\end{eqnarray}
The equations (\ref{eps1}) and (\ref{eos_1mn1}) show that the $A_1$ and $A_3$ four-vectors in the equations (\ref{W1}) and (\ref{W3}) can be considered as the boosted transverse polarizations, $\epsilon_{1,1}$, and $\epsilon_{1,-1}$, respectively. This observation, together with eq.~(\ref{divA_1})  qualifies  $A_1$ and $A_3$ as building blocks in the construction of the massless Rarita-Schwinger four-vector spinors in Section 5 below. In coordinate space, the vectors in (\ref{eps1})-(\ref{epszero}) take the shape of the pseudo-spherical harmonics via the relation, 
${\vec r}\cdot {\vec\epsilon}_{1}=Y_1^m(\cosh \rho)e^{im\varphi}$,  with ${\vec r}=(\sinh\rho\cos\varphi, \sinh\rho\sin\varphi,\cosh\rho)$ being a vector in an $(1+2)$ dimensional pseudo-Euclidean  position space. The pseudo-spherical harmonics are the eigenfunctions of ${\mathcal C}$ in its representation as a differential operator \cite{KimNoz}. \\

\noindent
Finally,  the equation (\ref{W2_lngt}) shows that for $p^2=0$, when $(E+p_z)^2-p_x^2-p_y^2=2(E+p_z)p_z$, the $(-iA_2)={ e}_{0,0}\equiv e_z$ four-vector is constant, which implies  that the Lorentz transformations are represented on it by the identity element. Put another way, this means that $(-iA_2)={e}_z$ transforms according to a trivial representation. In this way the massless $(1/2,1/2)$ representation of the Lorentz group, in contrast to the massive one \cite{AK}, \cite{ADK}, is not irreducible but  
splits into a trivial representation, i.e. an $so(1,2)$ singlet, 
defined by  the constant vector, $(-iA_2)=e_z$, coinciding with the unit vector of the $z$ axis, on one side, and an $so(1,2)$ triplet constituted by the two vectors, $(-iA_1)=\epsilon_{1,1}$, and  $(-iA_3)=\epsilon_{1,-1}$, of negative norms, and 
time-like vector $(-iA_4)=\epsilon_{1,0}\equiv n$ of a positive norm, on the other. 
In this fashion, within the scheme worked out here,  the purely space-like  polarization ${\hat e}_z$, transversal to $\epsilon_{1,1}$ and $\epsilon_{1,-1}$,  drops out of the kinematics of real massless spin-$1$ gauge fields, such as photons and gluons. The major conclusion of the current section is that  massless four vectors in the Weyl spinor approach are treated as massive vectors in a $(1+2)$ dimensional space time, with the $z$ component of the 
embedding $(1+3)$ dimensional space-time taking the role of ``mass,'' an observation known as well from light-cone field theories\footnote{The authors thank an anonymous referee of this article for bringing the cited articles to their attention.} \cite{Brodsky, Dirac49}, and from anyon physics \cite{Plyushchay3}. \\

\noindent In the subsequent section we shall compare the framework  of the present study with Wigner's little group approach to massless particles.

\section{Classification of the massless four-vectors by the helicity  quantum number and Wigner's little group method}
A brief comparison of the $SL(2,\C)$ approach  to massless states employed here with Wigner's induced representation method based on $E(2)$ as the little group is in order. 
The Euclidean group $E(2)$ is   defined as the semi-direct, $\rtimes$, product of the (Abelian) translation group on a plane, ${\mathcal T}_2$, with  the $SO(2)$ group of rotations, i.e. $E(2)={\mathcal T}_2\rtimes SO(2)$ and has been suggested by Wigner as a group whose representations induce in the Lorentz group the massless representations. 
The $E(2)$ generators can be chosen as the generator, $L_z$
 of rotation around the $z$ axis, and the two generators, $\Pi_x$ and $\Pi_y$,
 as the transverse Galilean boosts on the light cone \cite{Brodsky}, defined as \cite{Hitoshi}, \cite{Brink}
\begin{eqnarray}
\Pi_x=K_x+L_y,&& \Pi_y=K_y-L_x.
\end{eqnarray}
They commute with each other, $\left[\Pi_x,\Pi_y \right]=0$,
while their commutators with $L_z$ is given by,
\begin{eqnarray}
\left[\Pi_x,L_z\right]&=&-i\Pi_y, \quad \left[\Pi_x,L_z \right]=i\Pi_x.
\end{eqnarray}
Therefore, the three generators, $\Pi_x$, $\Pi_y$, and $L_z$  form the Galilean algebra of an Euclidean 2D space. 
Since $E(2)$ is non-compact, its unitary irreducible representations are infinite dimensional, while the finite dimensional ones are non-unitary. Here we are especially interested in Minkowski four-vectors, in whose space the $E(2)$ generators can be represented by the following matrices:
\begin{eqnarray}
\Pi_x=\left(
\begin{array}{cccc}
0&i&0&0\\
i&0&0&i\\
0&0&0&0\\
0&-i&0&0
\end{array}
\right), \quad \Pi_y=\left(\begin{array}{cccc}
0&0&i&0\\
0&0&0&0\\
i&0&0&i\\
0&0&-i&0
 \end{array}
\right), &\quad&
L_z=\left(\begin{array}{cccc}
0&0&0&0\\
0&0&-i&0\\
0&i&0&0\\
0&0&0&0
 \end{array}
\right).
\end{eqnarray}
It is easy to verify that the $\Pi_x$ and $\Pi_y$ operators have the property 
of annihilating the vector,
\begin{equation}
k^\mu= k \left( \begin{array}{c}
1\\
0\\
0\\
-1
\end{array}
\right)=k(\epsilon_{1,0}- e_z),
\label{kvector}
\end{equation}
the $L_{z}$ eigenstates associated to the zero eigenvalue that are the linear combination of $\epsilon_{1,0}\equiv n$ from (\ref{epszero}) and the unit vector along the $z$ axis, $ e_z$  
from (\ref{W2_lngt}).
{}Furthermore, the operators,
\begin{eqnarray}
\Pi_\pm &=&\Pi_x\pm i\Pi_y,
\end{eqnarray} 
have the following laddering  property,
\begin{eqnarray}
\Pi_x(\epsilon_{1,0}-e_z)=\Pi_y(\epsilon_{1,0}-e_z)=0,\quad 
\Pi_\pm \epsilon_{1,\mp 1}=\pm i\sqrt{2}(\epsilon_{1,0}-e_z).
\label{shift_oprtrs}
\end{eqnarray}
Therefore, the three vectors $\epsilon_{1,1}$, $\epsilon_{1,-1}$, and $k^\mu$ 
 behave as a massless helicity semi-triplet in the sense that the transverse $\epsilon_{1,\pm 1}$ can not be reached by laddering from $ (\epsilon_{1,0}-e_z)$, at variance to the algebra in (\ref{so12_lddrs}).
This happens because the $n$ vector in (\ref{epszero}) is not normalizable and describes a purely  time-like momentum, $p^2=1$, while  $k^\mu$ in (\ref{k_vector}) is not normalizable and describes a massless momentum, $k^2=0$. Gauge symmetry is known to remove the longitudinal degree of freedom both at the polarization vectors and quantum-state levels, and demands helicity conservation. \\

\noindent
 The property  of the transverse boost generators on the light cone to annihilate the  $k^\mu $  
vector is, however, not universal. Infinite dimensional unitary carrier spaces (infinite dimensional modules) can throughout be $\Pi_x$ and $\Pi_y$ eigenmodules with non-vanishing eigenvalues. In this case, by acting on such carrier spaces by the  $L_z$ operator,  infinite massless towers of equally spaced helicities  are created, known under the name of ``exotic'', ``continuous'', or ``infinite'' spin modules, \cite{Brink},  \cite{Repka} because the non-compact generators of the  boosts, present in $\Pi_x$ and $\Pi_y$, are not helicity conserving. According to \cite{BarutKleinert},  \cite{Plyushchay}, \cite{Plyushchay2}  Wigner's exotic state can be
associated  with infinite component Majorana fields, appearing as discrete  $D_\alpha^\pm$ series  of the $SL(2,R)$ group (it has  $E(2)$, and $SO(1,2)$ as subgroups) and considered in the limits of 
\begin{itemize}

\item  vanishing masses $m\to 0$, 

\item  infinitely growing spins, $s\to \infty$,  

\item  constant $ms\to \mu$ products,  with $\mu$ providing   new mass scales. 

\end{itemize}
As an example, the set of $n\ell$ hydrogen (H) atom states with a fixed  angular momentum value, $\ell$,  and a node-number $n$ growing from zero to infinity,  provide examples for  infinite component Majorana fields (with $n$ taking the part of ``spin'' index) as they can appear in composite systems. Taking into account the relative smallness  of the
binding energy of $(-13.6)$ eV of the H atom ground state relative to the electron mass of $m_e\approx 511$ eV, the infinite $n\ell$ towers in the H Atom spectrum could be considered as massless to a good approximation \cite{Itzhak}, 
thus  providing a reasonable illustrative example for an ``exotic'' massless state \cite{Plyushchay}.  However, no elementary particles following such patterns have been experimentally detected  so far in Nature,  although they are of interest to theories based in  higher than $(1+3)$ dimensions \cite{Bekaert}, \cite{Schuster}). In ordinary Minkowski space--time such fundamental ``exotic'' states have to be considered as non-physical, or, spurious. 
In restricting the $E(2)$ carrier spaces to such with vanishing $\Pi_x$ and $\Pi_y$ eigenvalues, the exotic states are excluded.
 As long as the maximal compact group of the little group is $SO(2)$ , which has only one-dimensional carrier spaces, each such space  is characterized by helicity. The fundamental $CPT$  symmetry then requires that each state of helicity $h$ has to be paired by an  opposite helicity, $(-h)$, meaning that
for each  spin $j$ one  finds a massless particle of  helicity $j$, and an antiparticle of helicity $(-j)$. \\

Finally, the connection between the $E(2)$ generators and the components of the 
Pauli-Lubanski vector can be established.
Indeed, in the case of $p_x = p_y = 0$, the Pauli-Lubanski vector is especially simple and reads,
\begin{equation}
W^\mu=
\frac{1}{2}k\left(
\epsilon^{\mu 0 \rho \sigma}M_{\rho \sigma} +
\epsilon^{\mu3\eta\tau }M_{\eta \tau}\right) .
\label{PL-msls}
\end{equation}
with $k$ being defined as before through $k^{\mu} = (k,0,0,k)$.
In executing the calculation, one verifies that this amounts to
\begin{equation}
\frac{W^0}{k}=\frac{W_z}{k}=L_z, \quad \frac{W_x}{k}=\Pi_x, \quad \frac{W_y}{k}=-\Pi_y.
\label{PLub}
\end{equation}
{}For this reason, some authors \cite{Das} prefer representing the $E(2)$ algebra in terms of the components of the Pauli-Lubanski vector given in (\ref{PL-msls}).

The main conclusion we draw from the present section is that Wigner's little group approach to massless four-vectors has the property to  
describe  genuine motion on the null-rays of the light cone.  In the Weyl spinor approach such motions are described through their projections on subspaces of the light cone, given by $SO(1,2)$ hyperboloids, thus re-phrasing the massless theory in terms of a massive one in a space time with one less spatial dimension. 
The Weyl spinor approach becomes comparable with the  $E(2)$ method upon giving up the $SO(1,2)$ irreducibility of the four-vectors in (\ref{so12_lddrs}), 
and of $e_z$ in (\ref{W2_lngt}), in which case the combination in (\ref{kvector}) becomes possible.

\section{Massless $(3/2,0)\oplus (0,3/2)$ as totally anti-symmetric  tensor-spinor}
Here we construct the tensor-spinors of the massless Rarita-Schwinger field from the four vectors built in the previous sections.
The massless Majorana Rarita-Schwinger particles of interest here  belong to the four-vector spinor carrier space of the Lorentz group, here denoted by 
$U^\mu$ and given by
\begin{eqnarray}
U^{(1)\mu}_{+}({\mathbf p})&:=& \epsilon_{1, 1}^\mu({\mathbf p})\otimes u^+ _M ({\mathbf p}),\\
U^{(1)\mu}_{-}({\mathbf p})&:=& \epsilon_{1, 1}^\mu({\mathbf p})\otimes u^- _M ({\mathbf p}),\\
U^{(2)\mu}_{+} ({\mathbf p})&:=& \epsilon_{1, -1}^\mu ({\mathbf p})\otimes u^+ _M({\mathbf p}),\\
U^{(2)\mu}_{-}({\mathbf p})&:=& \epsilon_{1, -1}^\mu({\mathbf p})\otimes u^- _M({\mathbf p}),
\end{eqnarray}
where $u_M^\pm({\mathbf p}) $  is one of the Majorana spinors in (\ref{Mjor_12}--\ref{Mjor_mn12}), while $\epsilon_{1,\pm 1}({\mathbf p})$ are the boosted transverse polarization vectors  from (\ref{eps1}--\ref{eos_1mn1}), i.e.
\begin{eqnarray}
\epsilon_{1,1}({\mathbf p})=A_1, \quad \epsilon_{1,-1}({\mathbf p})=A_3,
\end{eqnarray}
(see discussion after the equation (\ref{so12_lddrs})). In what follows the $ {\mathbf p}$ argument will be systematically suppressed for the sake of notational brevity. In taking into account the standard definition of the totally anti-symmetric Lorentz tensor of second rank,
\begin{eqnarray}
F^{\mu\nu}_1 &=&\epsilon^\nu_{1,1}p^\mu -p^\nu \epsilon^\mu_{1,1},\\
F^{\mu\nu}_2 &=&\epsilon^\nu_{1,-1}p^\mu -p^\nu \epsilon^\mu_{1,-1},
\label{FFmunu}
\end{eqnarray} 
which is divergenceless by virtue of equation (\ref{divA_1}), the following tensor-spinors are introduced:

\begin{eqnarray}
\left[ {\mathcal T}_\pm^{(1)}\right]^{\alpha\beta}&=& F_1^{\alpha\beta}\otimes u _M^\pm
=p^\alpha U^{\beta (1)} _\pm  -p^\beta  U^{\alpha (1)} _\pm ,
\label{T1}
\end{eqnarray}
and 
\begin{eqnarray}
\left[ {\mathcal T}_\pm ^{(2)}\right]^{\alpha\beta}&=&\ 
F_2^{\alpha\beta}\otimes u_M^\pm
=p^\alpha U^{\beta(2)}_\pm   -p^\beta  U^{\alpha (2)}_\pm .
\label{T_4}
\end{eqnarray}
Here,  we further suppressed the Majorana spinor index to avoid overloading the notation and used the notation  $U^{\mu(i)} _\pm $  from above for  the massless  Rarita-Schwinger four-vector-spinors.

According to (\ref{rdct}) all the ${\mathcal T}_\pm ^{(i)}$ spaces designed in this way are 24 dimensional, reducible,  and among their irreducible building blocks one encounters one  $(3/2,0)\oplus (0,3/2)$ irreducible tensor-spinor, that can be singled out by the pure spin-$3/2$  projector given in (\ref{F32_pryct}), constructed in \cite{AGK} as,

\begin{eqnarray}
\left[{\mathcal P}^{(3/2,0)}\right]^{ \alpha \beta} \,\, _{ \gamma \delta }\left[{\mathcal  T}_\pm ^{(i)}\right]^{ \gamma \delta }:=
\left[w^{(3/2,0)i}_\pm \right]^{\alpha \beta}.
\label{rdct_fnl}
\end{eqnarray}
According to \cite{Allcock}, the $(3/2,0)\oplus (0,3/2)$  carrier spaces of the  Lorentz group  are the only spin-$3/2$ representation spaces that respect Weinberg's theorem and  are  canonically quantizable.  These new spin-$3/2$ degrees of freedom satisfy the following relations,

\begin{eqnarray}
p_\alpha \left[w^{(3/2,0)i}_\pm \right]^{\alpha \beta}=0, \quad \gamma^\alpha\gamma^\beta \left[w^{(3/2,0)i}_\pm \right]^{\alpha \beta} =0.
\label{BDL_RRB}
\end{eqnarray}
As a reminder, the index $i=1,2$ refers to polarization vectors of helicity $(+1)$ and $(-1)$ respectively,  while the subscript $(\pm)$ specifies the helicity of the Majorana spinor, $(+1/2)$ versus $(-1/2)$.  The  tensor-spinors $\big[{\mathcal T}^{(1)}_-\big]^{\alpha \beta}$ and $\big[ {\mathcal T}^{(2)}_+\big]^{\alpha \beta}$   of helicities, $h=\pm 1/2$,  respectively, are expected to be 
ruled out  by gauge invariance in a similar way in which the longitudinal degrees of freedom of the four-vectors are removed. At the level of the states, these helicities are excluded by the subtle  Ward identities. Alternatively, as explained in \cite{Progress}  chiral  projectors based on the pseudo-scalar Casimir invariant, $G$, of the Lorentz algebra,

\begin{equation}
G_{AB}=\frac{1}{8}\epsilon_{\mu\nu}\,\,^{\alpha\beta}\left[S^{\mu\nu} \right]_{AB}\left[S_{\alpha\beta} \right]_{AB}, 
\end{equation}
where $S^{\mu\nu}$ are the generators of the homogeneous spin-Lorentz group, and $A, B=1,2,..., d$ are the indices of a $d$ dimensional representation, can be employed. 

On that basis, projector operators, here generically denoted by $\mathcal{P}_G^L$ and $\mathcal{P}_G^R$, respectively, can be constructed to have the property to split  the carrier space  under consideration into left $(L)$- and right $(R)$-handed degrees of freedom  according to, 
\begin{eqnarray}
\mathcal{P}_G ^R=\frac{1}{2}\frac{G+r}{r}, &\quad&
\mathcal{P}_G^L=-\frac{1}{2}\frac{G-r}{r}.
\label{PR_properties}
\end{eqnarray}
As an illustrative example we here bring the construction of such projectors on the single-spin  $(1/2,0)\oplus (0,1/2)$ space, where the spin-Lorentz group generators are $\frac{1}{2}\sigma_{\mu\nu}$. 
In this case, the eigenvalues of $G$ on the $(1/2, 0)\oplus (0,1/2)$ carrier space are given by
\begin{eqnarray}
G\, \phi^R&=&r\phi^R, \quad G \phi^L =-r \phi^L,
\label{G_Cas}\\
 r&=&i(K+1)M, \quad  K=1/2+0, \quad M=|1/2-0|,
\end{eqnarray}
where $\phi^R$ and $\phi^L$ are the respective right- and left-handed spinors in $(1/2,0)\oplus(0,1/2)$.
Since for the example under consideration the $G$ invariant calculates as 
$G=-\frac{3i}{4}\gamma^5$, the projectors emerge as
${\mathcal P}^L_G=(1-\gamma_5)/2$, and ${\mathcal P}_G^R=(1+\gamma_5)/2$, 
and  one recovers the well known  chiral spinors, $(u-v)/2$, and $(u+v)/2$. For  massless particles, the chirality projectors become  helicity  projectors.
{}For carrier spaces of multiple spins, $G$  correspondingly will have multiple eigenvalues  and the projectors will present themselves as products of single-helicity  projectors, a technique that was presented  for the very similar case of multiple spins in \cite{AGK}.

The most efficient tool for the explicit construction of a $\big[w^{(3/2,0)i}_\pm \big]^{\alpha \beta}$ tensor is using for $F_{\mu\nu}$ in (\ref{FFmunu})--(\ref{T_4}) the presentation suggested by Uhlenbeck and Laporte in \cite{UhLp} and based on spinorial indices. According to \cite{UhLp}, one defines a totally symmetric tensor of second rank, whose components are, 
$f_{11}=\chi_1\chi_1$, $f_{22}=\chi_2\chi_2$ and 
$f_{12}=f_{21}=(\chi_1\chi_2 +\chi_2\chi_1)/2$, and which  represents the  $(1,0)$ part from $(1,0)\oplus (0,1)$.  
Analogously, for $(0,1)$ one defines, $f^{\stackrel{\bullet}{ 1}\stackrel{\bullet}{ 1}}$, $f^{\stackrel{\bullet}{ 2}\stackrel{\bullet}{ 2}}$, $f^{\stackrel{\bullet}{ 1} \stackrel{\bullet}{ 2}}=f^{\stackrel{\bullet}{ 2}\stackrel{\bullet}{ 1}}$.  Then one introduces  the matrix
\begin{eqnarray}
f^\sigma\,_{\tau}=\left( \begin{array}{cc}
\chi^1\chi_1 & \chi^1\chi_2\\
\chi^2 \chi_1&\chi^2\chi_2
\end{array}\right)
&=&\left(\begin{array}{cc} 
{\mathcal F}^3&{\mathcal F}^1-i{\mathcal F}^2\\
{\mathcal F}^1+i{\mathcal F}^2&-{\mathcal F}^3
\end{array} \right),\quad {\mathcal F}^{i}={\mathbf E}^i-i{\mathbf H}^i, 
 \label{fstr_spnrs}
\end{eqnarray}
with ${\mathbf E}$ and ${\mathbf H}$ standing for the electric and magnetic fields. Then the $F^{\mu\nu}$ components  are expressed as

\begin{eqnarray}
 F^{0i}=-{\mathbf E}^i, &\quad& F^{ij}=-\epsilon^{ijk}{\mathbf H}^k.
\label{Fmunu_spnrs}
\end{eqnarray}
Taking, then, the direct product of $f^\sigma\,_\tau $ with the Majorana spinor $u_M^+$  in (\ref{Mjor_mn12}) will pick up from the latter the components suited for producing a totally symmetric tensor of third rank with spinor components, i.e. $f_{11}$ will pick up $\zeta_1$, $f_{22}$ will pick up $\zeta_2$, while $f_{12}$ can pick up either $\chi_1$ to form {\bf Sym}$\left(f_{12}\chi_1\right)$
, or $\chi_2$ to form {\bf Sym}$\left(f_{12}\chi_2\right)$. 
With $\chi^1$ and $\chi^2$ from (\ref{Weyl1}) and raising and lowering the spinorial indices according to the prescription in (\ref{C_spnrs})-(\ref{Cinvrs_cspnrs}) the following $f$ matrix is obtained:
\begin{equation}
f^\sigma\,_\tau =\left( 
\begin{array}{cc}
(E+p_z)(p_x+ip_y)& (p_x+ip_y)^2\\
-(E+p_z)^2& -(E+p_z)(p_x+ip_y)
\end{array}
\right)
\label{UL_1}
\end{equation}
This is a divergenceless tensor in accordance with

\begin{equation}
\left(\begin{array}{cc}
E+p_z& p_x-ip_y\\
p_x+ip_y& E-p_z
 \end{array}
\right)^T\left( 
\begin{array}{cc}
(E+p_z)(p_x+ip_y)& (p_x+ip_y)^2\\
-(E+p_z)^2& -(E+p_z)(p_x+ip_y)
\end{array}
\right)=0,
\label{UL_2}
\end{equation}
and in agreement with (\ref{BDL_RRB}). Now the totally symmetrized direct product,
\begin{equation}
\left( \frac{3}{2},0\right):\qquad  
\mbox{\bf Sym} \left( f \otimes \frac{1+\gamma_5}{2}u_M^+\right)=\Xi_{\alpha\beta\gamma}, 
\end{equation}
provides a representation of  the $(3/2,0)$ part of a $w^{(3/2,0)i}_+$ tensor spinor for $i=1$. The $\Xi_{\alpha\beta\gamma}$  tensor-spinor has four independent components, 
$\chi_1\chi_1\chi_1$, $\chi_2\chi_2\chi_2$, Sym$(\chi_1\chi_1\chi_2)$, and Sym$(\chi_2\chi_2\chi_1)$ from which only the first one survives in the $p_x=p_y=0$ gauge kinematics, as it should be. 
Notice that at the level of the spinorial indices it is the symmetrization procedure that plays the same role as the projector ${\mathcal P}^{(3/2,0)}$, given in (\ref{F32_pryct}), plays at the level of the Lorentz indices.  
In now recalling that the Maxwell equations are obtained from,

\begin{eqnarray}
i\left(
\begin{array}{cc}
\partial_0+\partial_3 & \partial_1-i\partial_2\\
\partial_1+i\partial_2 &\partial_0-\partial_3
\end{array}\right)
\left(  \begin{array}{cc}
{\mathbf E}^3-i{\mathbf H}^3 & ({\mathbf E}^1-i{\mathbf H}^1)-i({\mathbf E}^2-i
{\mathbf H}^2)\\
({\mathbf E}^1-i{\mathbf H}^1)+i({\mathbf E}^2-i{\mathbf H}^2)&-{\mathbf E}^3+i
{\mathbf H}^3 
\end{array}
\right)=0,
\label{laws}
\end{eqnarray}
as
\begin{equation}
\frac{\partial {\mathbf E}^3}{\partial t} -(\mbox{rot}{\mathbf H})^3
+\mbox{div}{\mathbf E}=0,
\end{equation}
and so on, one becomes aware how by working out the divergence of $\Xi_{\alpha\beta \gamma}$ upon its transformation to coordinate space, field equations for the gravitino of Maxwellian type could be encountered.

The corresponding classical Lagrangian and using generic notation could now be designed in parallel to (\ref{lr_ginvt_lgr}) as
\begin{eqnarray}
{\mathcal L}^{RS}_0&=&-\frac{1}{4}\left[w^{(3/2,0)}\right]^{\mu\nu}\left[
w^{(3/2,0)}\right]_{\mu\nu}.
\label{RS_lgr}
\end{eqnarray}
As an alternative, the massless limit of a Lagrangian for the $\Xi_{\alpha\beta\gamma}$ tensor,  transformed to coordinate space and there denoted by $\psi_{\alpha\beta \gamma}$,  can be considered along the lines of \cite{Srednicki}, \cite{Victor} and given by
 \begin{eqnarray}
{\mathcal L}&=& a\partial_\nu 
\psi^\dagger_{\stackrel{\bullet}{\alpha}\stackrel{\bullet}{\beta}\stackrel{\bullet}{\gamma}}
{\overline \sigma}^{\nu \stackrel{\bullet }{\alpha}\alpha}
{\overline \sigma}^{\mu\stackrel{\bullet }{\beta}\beta}
{\overline \sigma}^{\rho \stackrel{\bullet }{\gamma}\gamma}
\partial_\mu\partial_\rho \psi_{\alpha\beta\gamma} +
b \partial^\mu \psi^{\alpha\beta\gamma}\partial_\mu \psi_{\alpha\beta\gamma}+\nonumber\\
&+&\mbox{Terms\,\, of\,\, lower\,\, order\,\, in\,\, the\,\, derivatives}\nonumber\\
&+&m^2\psi^{\alpha\beta\gamma}\psi_{\alpha\beta\gamma}|_{m\to 0} +
\mbox{Hermitean \,\, conjugated\,\, terms}.
\label{spinor_Lagr} 
\end{eqnarray}
Here, $\overline{\sigma}^{\mu \stackrel{\bullet}{\alpha}\alpha } =\left( \sigma_0, -\vec {\sigma}\right) $,  $\vec \sigma =\left(\sigma_x,\sigma_y,\sigma_z\right)$, and 
$\sigma_x$, $\sigma_y$ and $\sigma_z$ are
the Pauli matrices, and $a$ and $b$ are constant parameters. In \cite{Victor} the case of a tensor-spinor of second rank has been worked out  along the lines of (\ref{spinor_Lagr}) by means of the canonical constraint Hamiltonian quantization procedure and there it could be shown that for $a+b=1$  the Hamiltonian is free from  negative energy solutions  and presents itself  diagonal in the particle creation and annihilation operators. 

Compared to the standard $(j,0)\oplus(0,j)\sim \Phi_B$ column-vector field description,
\begin{eqnarray}
\left(i^{2J}\left[
\gamma_{\mu_1\mu_2...\mu_{2j}} \right]_{AB}\partial^{\mu_1} \partial^{\mu_2}....
\partial^{\mu_{2j} } -m^{2J}\delta_{AB} \vert_{m\to 0}\right)\Phi_B&=&0, \quad A,B=1,..., 2(2j+1),
\label{Joos_Wnbrg}
\end{eqnarray}
with $\gamma_{\mu_1\mu_2...\mu_{2j}}$ standing for the Joos-Weinberg matrices,
Lagrangians of the type given in (\ref{spinor_Lagr}) have the advantage that the corresponding equations contain many more terms and thus  provide more chances of avoiding instabilities of the Hamiltonian through favorable  cancellations in the calculations of Dirac brackets. Work on this  program, both for massless and massive fields, has been initiated  in 
\cite{Victor}, and will be continued elsewhere. We expect massless pure spin 
$(j,0)\oplus (0,j)$ fields to become accessible to  canonical constraint Hamiltonian quantization in terms of Lagrangians of the type in (\ref{spinor_Lagr}). 
The spin-$3/2$ case considered here by us is supposed to describe a gauge field as is the gravition particle appearing in super-symmetric theories. 
For the sake of being employed as a field mediating gravitational interaction,
i.e. as a gauge field in supergravity, one will have  to replace in (\ref{spinor_Lagr}) the ordinary derivatives  by covariant ones which contain the spin connection, $\omega_\mu^{ab}=e^b\,_\nu \nabla_\mu e^{a \nu}$, where $e_\mu ^a$ is the vierbein. Then one could move on to elaborate schemes for couplings to matter fields. The aimed simplification of the quantization procedure presented above is expected to possibly facilitate some of the aspects of the  quantization procedure of  gravitational interactions, a topic for further studies.

\section{Conclusions}
In this work we presented the explicit construction of the massless Rarita-Schwinger four-vector spinors and the related anti-symmetric tensor-spinors of pure spin-$3/2$, framed in the formulation of the free wave equation given in (\ref{BDL_RRB}) that is distinct from the  one appearing in the  Rarita-Schwinger framework \cite{Sorokin}. The realization of this construction became possible by virtue of our knowledge of the momentum independent Lorentz projector in (\ref{F32_pryct}), earlier derived in \cite{AGK}, which served as a carrier space reduction tool. The wave equation in (\ref{BDL_RRB}) satisfied by this tensor-spinor seems reasonable in so far as it parallels the equation satisfied by the electromagnetic field tensor for the case of free photons,  $\partial^\mu F_{\mu\nu}=0$. We also discussed the relation between the Weyl spinor approach to massless four-vectors and Wigner’s little group method. In effect, we have shown that the Rarita-Schwinger approach to the massless gravitino is not unique and have provided a different option, which in our opinion bears the potential 
of being extendible to other high spins.  In future work we shall examine the quantisation of this approach and
the coupling to an electromagnetic field and compare the results to the more traditional formalism.

\subsection*{Acknowledgements}
The authors are glad to thank Christian Schubert for encouragement and many useful discussions.

\end{document}